\documentclass[11pt]{article}
\pdfoutput=1
 \usepackage{mciteplus}
 \usepackage{tikz}
 \usepackage{color}
 \usepackage{xcolor}
 \usepackage{comment}
 \definecolor{darkblue}{rgb}{0.1,0.1,.7}
 \usepackage[colorlinks, linkcolor=darkblue, citecolor=darkblue, urlcolor=darkblue, linktocpage,hyperfootnotes=false]{hyperref} 
\usepackage{epsfig}
\usepackage{graphicx}
\usepackage{cite}
\usepackage{amsfonts}
\usepackage{amssymb}
\usepackage{bm}
\usepackage{latexsym}
\usepackage{mathtools}
\usepackage{amsmath}
\numberwithin{equation}{section}
\def\bq{\begin{quote}}
\def\eq{\end{quote}}
 at 10truept

\newcommand{\cala}{{\cal A}}
\newcommand{\calC}{{\cal C}}
\newcommand{\calo}{{\cal O}}
\newcommand{\calh}{{\cal H}}

\newcommand{\call}{{\cal L}}
\newcommand{\caln}{{\cal N}}

\newcommand{\calq}{{\cal Q}}
\newcommand{\calx}{{\cal X}}

\newcommand{\beq}{\begin{equation}}
\newcommand{\eeq}{\end{equation}}
\newcommand{\beqa}{\begin{eqnarray}}
\newcommand{\eeqa}{\end{eqnarray}}
\newcommand{\bea}{\begin{eqnarray}}
\newcommand{\eea}{\end{eqnarray}}

\newcommand{\hf}{\frac{1}{2}}

\def\roughly#1{\raise.3ex\hbox{$#1$\kern-.75em\lower1ex\hbox{$\sim$}}}

\newcommand{\tA}{{\tilde A}}
\newcommand{\tB}{{\tilde B}}
\begin{document}

\thispagestyle{empty}
\begin{titlepage}
  \bigskip

  \bigskip\bigskip

  \bigskip

\begin{center}
{\Large \bf {Comparing models for a unitary black hole S-matrix
}}
    \bigskip
\bigskip
\end{center}

  \begin{center}

 \rm {Steven B. Giddings\footnote{\texttt{giddings@ucsb.edu}} }
  \bigskip \rm
\bigskip

{Department of Physics, University of California, Santa Barbara, CA 93106, USA}  \\
\rm

  \bigskip \rm
\bigskip
 
\rm

\bigskip
\bigskip

  \end{center}

\vspace{3cm}
  \begin{abstract}
This paper compares features, challenges, and puzzles of different models for a unitary black hole S-matrix, focussing on both recent nonisometric models, as well as ``nonviolent unitarization," which is based on new quantum interactions of a black hole.  As a foundation for comparison, the description of real-time Hawking evolution is first overviewed, including leading effects of gravitational dressing and backreaction.
Connection is then made to qubit models for evolution, and some technology is outlined to facilitate their description. 
Important features of both nonisometric models and nonviolent unitarization are investigated in qubit models, which illustrate essential differences between the respective approaches.  The nonisometric models present puzzles for understanding evolution of internal outgoing excitations, which can be excited by interactions such as particle decay.  Qubit models for nonviolent unitarization are further developed, and nicely illustrate aspects of that approach.
Some remaining questions in generalizing to more complete models for evolution are discussed.  

 \medskip
  \noindent
  \end{abstract}
\bigskip \bigskip \bigskip 

  \end{titlepage}

\section{Introduction}

Consider a cavity with reflecting walls, and a tiny hole through them through which we inject a very high energy photon in a pure state.  This photon will generically create additional particles, which evolve towards an approximately thermal distribution inside the cavity.  Over a long time, particles will also  find the small hole, and leak out.  The total state of particles will be pure, but the escaping particles will be entangled with those inside the cavity, and so a density matrix for them will be mixed and initially will have an increasing von Neumann entropy.

After a very long time, all the particles will have leaked out of the cavity, which will now be in its vacuum, and the state of the radiation outside will correspondingly be a pure state if one accounts for all the escaped particles.  

If one didn't understand the fundamental dynamics but had seen proposals about implementing a final state boundary condition in quantum mechanics\cite{ABL,Griffiths:1984rx,Gell-Mann:1991kdm},
it might be tempting to propose a new rule,  the final state proposal for cavities: that ordinary unitary evolution is modified so that after a long time the fundamental evolution  projects on the interior ground state.  

However, this is obviously unnecessary, and doesn't represent the correct fundamental dynamics. It is the small interactions between cavity modes and modes of the exterior, arising from the hole, combined with the infinite phase space of the exterior of the cavity, that accomplishes the same thing.  No new rule is needed.

Black hole (BH) formation and evaporation presents us with a seemingly very similar situation, where an initially simple, even pure, state can form a black hole, in whose interior we expect a complicated state to result.  Particles leak out, and ultimately we expect the black hole to disappear, effectively returning its interior to vacuum and the outside state to the same purity to avoid other contradictions arising from information loss\cite{Hawk-incoh,BPS} or remnants\cite{CaWi,Pres,WABHIP,Susstrouble}.  

What is different is that a description of BH evolution based on that of local quantum field theory (LQFT) on a semiclassical background, while allowing particles to escape\cite{Hawk}, does not allow entanglement (which can be used to track quantum information) to escape: its escape is forbidden by locality with respect to the semiclassical background geometry.  There are different possible reactions to this.

One is to propose that a fundamentally new principle like  the final state proposal\cite{ABL,Griffiths:1984rx,Gell-Mann:1991kdm} now comes into play\cite{HoMa}, and enforces the ultimate triviality of the BH internal state, while implementing unitarity of the full S-matrix.  A newer variant on this theme is discussed in the models of \cite{AEHPV}, which also successively incorporates final state projections, fundamentally modifying the evolution.

Another is to explore a resolution  analogous to evolution of conventional quantum systems like our cavity, arising from small interactions between BH states and those of the surroundings.  Of course, the presence of such interactions still appears rather radical, since such interactions would appear to violate the principle of locality -- specifically the requirement that information propagates inside the lightcone -- with respect to the background semiclassical geometry.  Indeed, if such interactions are present, they may well be evidence of some more fundamentally quantum properties of spacetime itself, and so ultimately are plausibly also associated with a modification of fundamental principles.

Either way, we confront the apparent need for new fundamental principles.  This paper will further compare and contrast 
these two scenarios,  treated as modifications to the more concretely understood basic evolution described by LQFT, updating Hawking\cite{Hawk} and many intervening works.

One can consider other possibilities as well; our thinking about them can be organized by considering a ``black hole theorem\cite{BHthm}."  A first question is whether a BH can be thought of as a quantum subsystem of a larger system.  There has been progress on seeing that this is possible, at least to a good approximation\cite{DoGi4,SGsplit,SGsub}, despite arguments to the contrary\cite{CGPR}, but questions remain on the fundamental structural question of how to define subsystems of quantum gravitational systems, and these conceivably present a different out.  Or, perhaps other dynamics, {\it e.g.} that of fuzzballs\cite{Fuzzrev}, could modify that assumption.  But, assuming that BHs {\it do} effectively behave as subsystems, and that we wish to avoid the apparent contradictions of information loss or microscopic remnants, it appears that such a theorem then leaves one with the choices of modification of quantum evolution, or small interactions, or scenarios that are an even greater departure from the tried and tested physics of LQFT, and which also have their own problems.

In order to better understand the nature of modifications to LQFT evolution, this paper will first give in section two a brief overview of aspects of a more detailed description of such real time evolution, including a leading order treatment of quantized general relativity (QGR), incorporating the leading effects of the quantum gravitational field:  dressing and backreaction.  This provides a basis for comparison of modifications to such evolution, particularly if such modifications can be treated as corrections to such evolution that are in some sense ``small."\footnote{Such modifications must of course have a significant effect -- the $\calo(1)$ departure of the Page curve\cite{Pageone,Pagetwo} from the rising entropy of Hawking radiation, as also emphasized in \cite{Mathinforev}.} The next section will discuss some basic considerations regarding modifying that evolution.  Section four then investigates qubit models for Hawking evolution and for its modification either via nonisometric evolution, or unitarizing interactions.  In the process it develops some useful technology for describing qubit models, and  investigates the real-time evolution of the nonisometric models as well as potential issues that they raise.  It also
 further develops nonviolent models for unitarizing interactions and dynamics.  Section five concludes with discussion of some questions associated with extension of such models to a more complete description of evolution, and of further comparison of models in this context.

\section{Evolution in perturbative LQFT/QGR: overview}
\label{QFTsec}

This section will briefly review the  evolving quantum state of a black hole, as described within perturbative LQFT/QGR.  Many features of this have been understood for decades, though aspects of a more complete description, including the role of gravitational dressing/backreaction, have been provided recently in \cite{GiPe1,GiPe2}.

We consider perturbations about a background metric $g_{\mu\nu}$, which we take to describe a BH that formed from collapse.  At later times, the classical solution settles down to a stationary black hole; we focus on the simplest  static case, with radius $R$. If we account for Hawking radiation, the mass decreases at the rate $dM/dt \sim 1/R^2$, which can be accounted for by including backreaction of the expectation value of the leading quantum stress tensor in determining $g_{\mu\nu}$.\footnote{This is clearest in the example of two-dimensional dilaton gravity models\cite{CGHS,SGTrieste}.}  
However, since the fractional mass decrease during the typical emission time $R$ of a quantum is $\Delta M/M\sim 1/MR\sim 1/S_{BH}$, with $S_{BH}$ the Bekenstein-Hawking entropy, this metric is to very good approximation static over many such emission times, during the period after it settles down from initial formation, and before its final demise.  
We consider quantum perturbations of both infalling and outgoing excitations.
%

Gauge fixing is needed to describe the evolving quantum state, including  matter and metric perturbations, of the BH.  Specifically, in QGR this gauge fixing can be accomplished by providing a family of time slices, and spatial coordinates on these slices, described through functions
\beq\label{slices}
x^\mu =\calx^\mu(t, x^i)\ 
\eeq
parameterizing these slices in general coordinates $x^\mu$.  Such slices, and evolution on them, are further discussed in \cite{SEHS,SE2d}\cite{GiPe1}.
It is simplest to choose these slices to respect the spherical symmetry of the background metric.  Using the standard radial coordinate and an ingoing null coordinate $x^+$, these simplify to
\beq\label{radslices}
x^+=\calx^+(t,x)\quad ,\quad r=\calx^r(t,x)\ .
\eeq
On the long time scales over which the static approximation holds, the metric is approximately
\beq\label{Schw}
ds^2=-f(r) dx^{+2} + 2 dx^+ dr + r^2 d\Omega_{D-2}^2\ ,
\eeq
with
\beq
f= 1-\left(\frac{R}{r}\right)^{D-3}\ .
\eeq
It can be advantageous to take the slices to respect the time-translation symmetry $x^+\rightarrow x^+ + a$, and so take the simplified stationary 
form\cite{NVU}\cite{SEHS}\cite{GiPe1}
\beq\label{statslices}
x^+= t+s(x)\quad,\quad r=r(x)\ .
\eeq

The Arnowitt-Deser-Misner (ADM) parameterization\cite{ADM}
\beq
ds^2= -N^2 dt^2 + q_{ij}(dx^i + N^i dt)(dx^j + N^j dt)
\eeq
gives the general form of a time-sliced metric, with normal $n^\mu=(1,-N^i)/N$ to the slices.  We will describe the evolving quantum state of the BH using these variables.  We consider  the small $G$ limit while keeping $G^{1/(D-2)}M$ large to maintain a large black hole in Planck units,
and work first to zeroth order in Newton's constant $G$, and then include perturbative corrections in $G$.

\subsection{Vanishing $G$ limit}
\label{Gez}

 For simplicity we consider scalar matter with lagrangian
\beq
S=-\hf \int d^D x \sqrt{|g|} (\nabla\phi)^2 \ ;
\eeq
we expect the treatment to generalize to matter with higher spin.  In the vanishing $G$ limit we neglect the backreaction of matter fluctuations on the metric.  The conjugate momentum to $\phi$ is 
\beq
\pi= \frac{1}{\sqrt q}\frac{\delta S}{\delta \dot\phi} =  n^\mu\partial_\mu \phi\ ,
\eeq
in terms of which the canonical form of the action becomes
\beq
S=\int dt d^{D-1}x \sqrt q \left(\pi \dot\phi  - \calh\right)\ ,
\eeq
with hamiltonian
\beq\label{matH}
H =  \int d^{D-1} x \sqrt q \calh= \int d^{D-1} x \sqrt q \left[ \frac{N}{2}\left(\pi^2+q^{ij}\partial_i\phi \partial_j\phi\right) + \pi N^i \partial_i\phi \right] \ .
\eeq 

The form of the classical solutions for $\phi$ plays an important role in  quantization.  Working in the static background \eqref{Schw}, these may be separated into two orthogonal spaces\cite{GiPe1}.  The first are the ``in" solutions, $\Phi_{\tilde A}(t,x^i)$.  These are defined such that as $t\rightarrow-\infty$, well-localized wavepacket superpositions 
are incoming and supported only in the asymptotic large-$r$ region.  In the future, they have components that are both outgoing, and that travel towards $r=0$.  (In the 2d case of conformal matter, these are left movers.)  The other ``up" space consists of modes $\Phi_A(t,x^i)$ such that as $t\rightarrow-\infty$ wavepackets become increasingly localized at $r=R$, and outgoing;\footnote{For $r<R$, outgoing means asymptotically tracks the ``outward" null direction.} in the actual limit they become singular at $r=R$.  In the future, they also have both outgoing and infalling components.  (In the 2d conformal case, these are the right movers.)  A basis for these modes can be chosen that is regular at the horizon, but if one instead chooses an energy eigenbasis with respect to a stationary slicing, that basis is singular at $r=R$ and can be furthermore separated into orthogonal subspaces of modes with outgoing component near $r=R$ restricted either to $r<R$ or $r>R$, respectively\cite{GiPe1}.

Orthogonality of these modes is with respect to the inner product
\beq\label{InPr}
(\Phi_1,\Phi_2) = i \int d^{D-1} x \sqrt{q} n^\mu \Phi_1^*\overleftrightarrow{\partial_\mu} \Phi_2\ ,
\eeq
which is conserved by the equations of motion.  This means it can for example be evaluated at $t\rightarrow-\infty$, demonstrating orthogonality of the different subspaces as defined above.  Given a time slice, a solution $\Phi(t,x^i)$ corresponds to Cauchy data $\gamma(x^i)=(\phi(x^i),\pi(x^i))$ on that slice, and \eqref{InPr} induces the inner product
\beq\label{CauPr}
(\gamma_1,\gamma_2)= i\int d^{D-1}x \sqrt{q}(\phi_1^*\pi_2-\pi_1^*\phi_2)
\eeq
on that Cauchy data, which is also conserved.  If we consider the fully extended BH solution \eqref{Schw}, Cauchy slices can be defined by taking the interior portion of the slices to asymptote to $x^+=-\infty$, for example as $r\rightarrow R_n<R$; otherwise one must confront the question of the form of the dynamics near $r=0$.

Description of the evolving wavefunction begins with the canonical commutation relations 
\beq\label{CCR}
[\pi(x^i, t), \phi(x^{i\prime}, t)]=-i \frac{\delta^{D-1}(x-x')}{\sqrt{q}}\ .
\eeq
We consider quantization with respect to a time slicing \eqref{slices} or \eqref{radslices}.\footnote{Due to additional subtleties of time-dependent metrics\cite{ToVa,CMOV,CFMM,AgAs,MuOe}, we typically also restrict to stationary slicings \eqref{statslices}.}  Assume that there is a separation of the modes into complex bases $(\Phi_A,\Phi^*_A)$, $(\Phi_{\tilde A},\Phi^*_{\tilde A})$, with conjugates orthogonal,\footnote{This can be described in terms of a complex structure on the space of solutions.} and such that the modes $\Phi_A$, $\Phi_{\tilde A}$ are positive norm under \eqref{InPr}.  Then, in terms of the corresponding Cauchy data on some fiducial $t=$constant slice, expand
\bea\label{piphiexps}
\phi(x^i) &=& \sum \big[a_A \phi_A(x^i)+a_A^{\dagger} \phi_A^*(x^i)+ a_\tA \phi_\tA(x^i)+a_\tA^{\dagger} \phi_\tA^*(x^i)\big]\  ,\cr
\pi(x^i) &=& \sum \big[a_A\pi_A(x^i)+a_A^{\dagger} \pi_A^*(x^i)+a_\tA\pi_\tA(x^i)+a_\tA^{\dagger} \pi_\tA^*(x^i)\big]\ .
\eea
If the mode bases are taken to be orthonormal, $(\gamma_A,\gamma_B)=(\gamma_\tA,\gamma_\tB)=\delta_{AB}$, the canonical commutators become
\beq
[a_A, a_B^{\dagger}] = [a_\tA, a_\tB^{\dagger}] =\delta_{AB}\quad ,\quad  [a_A, a_B]=[a_A^{\dagger}, a_B^{\dagger}] = [a_\tA, a_\tB]=[a_\tA^{\dagger}, a_\tB^{\dagger}] =0\ ,
\eeq
along with vanishing commutators between in and up operators, 
and Fock space states may be constructed by acting with raising operators on the vacuum $|0\rangle$ annihilated by $a_A$ and $a_\tA$.  In occupation number basis, these take the form $|\{n_A\},\{n_\tA\}\rangle$.

For a ``gauge-fixing" specified by a  given slicing, determining $(N, N^i, q_{ij})$, the hamiltonian \eqref{matH} can be rewritten in terms of the annihilation and creation operators by substitution of \eqref{piphiexps}.\footnote{Different such descriptions of the state arising from different gauge fixings are then expected to be related by unitarity transformations, with an analogy to change of picture.  One approach to arguing for such equivalence is expected to be based on the Hadamard behavior of the state\cite{KaWa}.} This gives an expression
\beq\label{Hama}
H = \sum_{A,A'} \left(A_{AA'} a^\dagger_A a_{A'} + B_{AA'} a^\dagger_A a^\dagger_{A'} + c.c. \right) + (A\rightarrow \tilde A)\ ,
\eeq
where $c.c.$ denotes conjugation that doesn't change operator ordering, and there is no mixing between $A$ and $\tilde A$ modes.  The coefficients $A_{AA'}$, $B_{AA'}$, $A_{\tA\tA'}$, $B_{\tA\tA'}$ may be expressed in terms of the mode basis and the metric coefficients, with explicit expressions given in \cite{SEHS,SE2d}\cite{GiPe1}.  
A specified initial state is then evolved by
\beq\label{Uev}
U(t_2,t_1)= e^{-i\int_{t_1}^{t_2} dt H}\ .
\eeq

In addition to standard propagation described by $a^\dagger a$ terms, $H$ creates paired excitations through the $a^\dagger a^\dagger$ terms.  One way to understand the detailed structure of this pairing is to rewrite the state in a basis of energy eigenmodes, as described above.  
Then the hamiltonian takes the form\cite{GiPe1}
\beq\label{Eham}
H= \sum_{lm} \int \frac{d\omega}{4 \pi \omega} \omega (b_{\omega lm}^{\dagger}b_{\omega lm} - \hat b_{\omega lm}^{\dagger} \hat b_{\omega lm} + \tilde b_{\omega lm}^{\dagger}\tilde b_{\omega lm}  )  \ ,
\eeq 
up to a normal-ordering constant, with the annihilation operators $b_{\omega lm}$, $\hat b_{\omega lm}$, $\tilde b_{\omega lm}$ corresponding to external up modes, internal up modes, and in modes, respectively.
Such a basis is singular, and makes contact with Hawking's original arguments\cite{Hawk}.  In such a basis, a regular state such as $|0\rangle$ has pairing between excitations on either side of the horizon, with thermal coefficients; at high wavenumbers such a regular state has behavior (specializing to $D=4$)
\beq\label{stateform}
|\psi\rangle \sim\sum_{\{n_{\omega lm}\}} e^{-2\pi R\int d\omega \omega n_{\omega lm} }| \{\hat n_{\omega lm}\} \rangle |\{n_{\omega lm}\}\rangle\ .
\eeq
The diagonalized hamiltonian \eqref{Eham} means its evolution is trivial, with inside modes having negative energies.  If one works in a wavepacket basis of these modes, one can follow an outside mode into the asymptotic region, where it can alternately be described in terms of a regular mode.  This mode is paired with internal excitations, through \eqref{stateform}; more discussion of this argument appears in \cite{GiPe1}.  All initially regular states are argued to have the same late time behavior; differences correspond to transitory excitations that fall deep into the BH or escape to infinity, and continuing long-time production of paired excitations then follows from the uniform vacuum-like structure of such states at short distances near $r=R$.

In principle, one should be able to see much of this structure directly in terms of a regular basis of modes.  Since localization plays an important role, we can imagine choosing bases $\Phi_A$, $\Phi_\tA$ that are approximately localized both in space and in wavenumber.  One such basis comes from window functions in frequency\cite{Hawk,GiNe}, although one may wish to optimize localization even further by using other bases, for example a wavelet basis.  In addition, with such a basis of localized modes (modulo some rapidly falling tails), one may separate the up modes $\Phi_A$ into sets $\{\Phi_a\}$, $\{\Phi_{\hat a}\}$ of modes dominantly localized in the regions $r>R$ and $r<R$ respectively.  In such a basis it is in practice much harder to evaluate the $A$ and $B$ coefficients in the hamiltonian, and the unitary evolution operator \eqref{Uev}, although in principle they can be calculated.  But due to the approximate correspondence of these wavepackets to those of the singular basis in regions well away from the horizon, we should approximately have the same entanglement structure as in \eqref{stateform}, with approximately thermal weights, when the outside modes are at $r\gg R$; the corresponding inside modes will then typically be deep inside the BH.  Of course, in the region $r\sim R$ this description of the state differs significantly -- in the regular basis, the state looks like the vacuum for modes very close to the horizon\cite{SEHS,SE2d}.

This discussion should also straightforwardly generalize to other types of matter, with spin.  The choice of the mode basis of course becomes a little more intricate, and one may also need to specify some appropriate gauge fixing.  But otherwise, one may similarly canonically quantize, and for a free field, find a hamiltonian with the same structure as \eqref{Hama}, resulting in a similar structure for the evolving quantum state, with particle production and internal/external pairing, and approximately thermal weights.  This includes the case of spin two perturbations, corresponding to fluctuations of the metric.

The approach described here also has the virtue of readily (in principle) generalizing to the case of interacting fields.  This is in contrast to conventional derivations of Hawking radiation.  There, modes are traced back via free evolution to the near-horizon region, in order to match onto the short-distance behavior there, and this procedure requires significant modification in an interacting theory.  In the present treatment using regular modes, interactions simply lead to higher-order terms in \eqref{Hama}, whose contribution to evolution through \eqref{Uev} may in principle be followed.  Of course this is no simpler than the description of any other interacting QFT system, but does have the virtue of being placed on a foundation involving regular rather than singular modes.

\subsection{Leading $\calo(G)$ corrections: dressing and backreaction}

While the full quantum description of the evolving gravitational state remains somewhat mysterious, one can apparently characterize the perturbative behavior of its small fluctuations, for example to leading order in $G$, working in QGR.  The starting point is the action
\beq\label{gravaction}
S=\int d^Dx \sqrt{|g|} \left( \frac{1}{16\pi G} R + \call_m\right) + S_\partial
\eeq
where $\call_m$ is the matter lagrangian and $S_\partial$ is a surface term.  
The conjugate momenta to $q_{ij}$ are
\beq
P^{ij}=  \frac{\delta S_g}{\delta \dot q_{ij}} = -\frac{\sqrt q}{16\pi G}\left(K^{ij}- q^{ij} K\right)\ ,
\eeq
given in terms of the extrinsic curvature
\beq\label{excu}
K_{ij}= \frac{1}{2N}\left(-\dot q_{ij} + D_iN_j + D_jN_i\right)\ 
\eeq
of the slices, with $D_i$ the covariant derivative with respect to $q$.  The conjugate momenta to $N$ and $N^i$ vanish, corresponding to their role as Lagrange multipliers for constraints.  These result in the canonical form for the gravitational part of the action
\beq\label{CGact}
S_g = \int d^D x \left( P^{ij}  \dot q_{ij} -  \calh_g\right)\ .
\eeq
The full hamiltonian may be written in two ways,\footnote{For more detailed expressions, see \cite{GiPe2}.} as
\beq\label{Hreln}
H = \int d^{D-1}x (\calh_g+ \calh_m) =  \int d^{D-1}x \left(N \calC_n + N^i \calC_i\right) + H_\partial\ .
\eeq
Here $\calh_g$ is quadratic in momenta and spatial derivatives of the metric variables.  In the second expression, $\calC_n$, $\calC_i$ are the constraints, 
\beq
\calC_\mu =\sqrt q \left( -\frac{G_{\mu\nu}}{8\pi G} + T_{\mu\nu} \right) n^\nu\ ,
\eeq 
with $\calC_n=n^\mu \calC_\mu$.   $H_\partial$ is a boundary term containing the ADM energy and momentum, and when the constraints are satisfied, $\calC_\mu=0$, the hamiltonian equals this boundary term.

There are different approaches to quantization; here we take that of ``gauge-invariant canonical quantization," which is closely related to covariant gauge fixing (breaking)\cite{GiPe2}.  The canonical commutators are taken to be
\beq\label{ccrq}
[P^{ij}(x,t),q_{kl}(x',t)] = -i\delta^i_{(k}\delta^j_{l)}  \delta^{D-1}(x-x')\ ,
\eeq
with $\delta^i_{(k}\delta^j_{l)}=(\delta^i_{k}\delta^j_{l} + \delta^i_l\delta^j_{k})/2$, and evolution is generated by the hamiltonian, as summarized by the Heisenberg equations,
\beq\label{HEOM}
\partial_t q_{ij} = i [H,q_{ij}]\quad ,\quad \partial_t P^{ij} = i [H,P^{ij}]\ .
\eeq
Since the constraints generate gauge transformations, gauge-invariant operators are those that commute with the constraints,
\beq\label{concomm}
[\calC_\mu(x),O]=0\ .
\eeq
However, the constraints should not annihilate physical states, as that would be inconsistent with the equations of motion \eqref{HEOM}.  Instead, the constraints are decomposed into pieces
 \beq\label{condec}
\calC_\mu(x)= \calC_\mu^+(x) + \calC_\mu^-(x)\ ,
\eeq
corresponding for example to positive and negative frequency contributions, and physical states are taken to obey
\beq\label{physS}
\calC_\mu^+(x)|\psi\rangle=0\ .
\eeq
Given a vacuum state satisfying \eqref{physS}, other physical states may then be constructed by acting with a gauge-invariant operator, $|\psi\rangle = O|0\rangle$.  With such states, evolution of matrix elements of gauge-invariant operators, 
\beq\label{MEevo}
\partial_t \langle \psi'|O |\psi\rangle = i\langle \psi'|[H,O]|\psi\rangle\ 
\eeq
is gauge invariant, {\it i.e.} independent of $N, N^i$.  

The gauge-invariance conditions \eqref{concomm}, \eqref{physS} may be solved perturbatively by constructing a gravitational dressing\cite{SGAlg,DoGi1,DoGi2,DoGi3,QGQF}\cite{DoGi4,SGsplit}.\footnote{Earlier work on this question includes \cite{Heem} and  \cite{KaLigrav}.   The first derived nontrivial commutators as arising from the constraints, but didn't give the dressed operators, and the second focussed on deriving {\it commuting} bulk operators.}  Defining $\kappa^2=32\pi G$, the metric variables are expanded about a background metric as
\beq
\tilde g_{\mu\nu} = g_{\mu\nu} + \kappa h_{\mu\nu}\quad  , \quad 
\tilde q_{ij}= q_{ij} + \kappa h_{ij}\quad ,\quad
\tilde P^{ij} = P^{ij} + \frac{p^{ij}}{\kappa}\ ,
\eeq
and the commutators \eqref{ccrq} become
\beq\label{Pcom}
[p^{ij}(x,t),h_{kl}(x',t)] = -i \delta^i_{(k}\delta^j_{l)} \delta^{D-1}(x-x')\ .
\eeq
The constraints may then be expanded in $\kappa$, as described in \cite{GiPe2}.  
For a given operator $O_0$ of the underlying field theory (here, of $\phi$), a gauge-invariant solution to \eqref{concomm}, to first order in $\kappa$, takes the form
\beq\label{invO}
O=e^{i\int d^{D-1}x \sqrt{q} V^\mu(x)T_{n\mu} } O_0 e^{-i\int d^{D-1}x \sqrt{q} V^\mu(x) T_{n\mu} }\ ;
\eeq
this may also be generalized to include dressing of gravitational perturbations by including a gravitational pseduo-stress tensor contribution to the stress tensor $T_{n\mu}$.  Here $V^\mu(x)$ are functionals of the metric perturbation.  Leading order gauge invariance follows if these functionals take the form
\bea\label{genv}
V^n(x)&=&\frac{\kappa}{2} \int d^{D-1}x' \left[ \check h_{ij}(x',x) p^{ij}(x') - \check p^{ij}(x',x) h_{ij}(x')\right]\ ,\cr
V^i(x)&=& \kappa \int d^{D-1}x' \left[ G^{ijk}(x',x)h_{jk}(x') + H^i_{jk}(x',x) p^{jk}(x')\right]\ 
\eea
where $\check h_{ij}$,  $\check p^{ij}$, $G^{ijk}$, $H^i_{jk}$ are Green function solutions of the linearized metric perturbation problem, detailed in \cite{GiPe2}.  These are not uniquely specified, in the absence of boundary conditions: to any such solution, we can add a solution of the homogeneous equations, corresponding to a freely-propagating gravitational wave.  This leads to a corresponding nonuniqueness in the gravitational dressing.

A matter state $|\psi\rangle_0$ (or one of perturbative gravitons) may be likewise dressed
\beq\label{GIstate}
|\psi\rangle = e^{i\int d^{D-1}x \sqrt{q} V^\mu(x)T_{n\mu} } |\psi\rangle_0\ ,
\eeq
and will solve \eqref{physS} if the underlying state $|\psi\rangle_0$ is also taken to solve the $\kappa=0$ version of equation \eqref{physS}.

The dressing in \eqref{GIstate} has the anticipated effect of creating the linearized quantum gravitational field arising from the stress energy of the underlying state.  Again, this is not unique since it can be augmented by freely propagating gravitational waves.  Examples of this non-uniqueness are provided in \cite{DoGi1}, where one may for example consider initial conditions where the gravitational field configuration is the analog of the Coulomb field, or is instead concentrated in a single gravitational line stretching to infinity.

Evolution of an initial dressed state \eqref{GIstate} will then be governed by the hamiltonian \eqref{Hreln}.  The leftmost form of this hamiltonian exhibits the terms responsible for evolving the metric perturbation and the matter perturbation, in the standard LQFT fashion.  For example, an initial linelike gravitational field like that just described will transition to an energetically favorable field like that of Coulomb, with gravitational radiation escaping to infinity.  This treatment of the leading gravitational field thus also describes the leading quantum backreaction of the underlying matter state, that is it accounts for the leading correction to the gravitational background $g_{\mu\nu}$ driven by the stress tensor of that matter state.

\subsection{General features of the perturbative quantum state}

Beginning with an initial state $|\psi(0)\rangle$, the perturbatively evolved $G=0$ Fock space state can be expanded in the general form 
\beq \label{matstate}
|\psi(t)\rangle = \sum_{\{n_a\}\{n_{\hat a}\}\{n_{\tilde A}\}} \cala\left({\{n_a\}\{n_{\hat a}\}\{n_{\tilde A}\}},t\right)|{\{n_a\}\{n_{\hat a}\}\{n_{\tilde A}\}}\rangle\ ;
\eeq
for example, by relating this to evolution of \eqref{stateform}, as described in \ref{Gez}, we find entanglement between the internal $a^\dagger_{\hat a}$ and external 
$a^\dagger_{a}$ excitations that grows with time.  This can be characterized by introducing an appropriate regulator, {\it e.g.} cutting off near horizon excitations with wavelengths $\ll R$, and tracing out internal excitations to find a density matrix with a growing von-Neuman entropy.\footnote{See {\it e.g.} \cite{GiNe} for a simple two-dimensional example.}  Continued growth of this during the BH evolution to a value comparable to the Bekenstein-Hawking entropy $S_{BH}$ is evidence of the missing information.

The leading perturbative correction to this arises from evolving the dressed version \eqref{GIstate} of the state \eqref{matstate} by the hamiltonian \eqref{Hreln}; as described, this evolution incorporates the gravitational constraints.
The gravitational dressing {\it does} introduce additional subtleties in giving a precise characterization of the BH subsystem and calculation of its entropy.
An open question is how to sharply characterize gravitational subsystems\cite{DoGi4,SGsplit,SGsub}, taking into account such effects at this and higher orders in $\kappa$.  It has even been argued that the dressing eliminates the missing information\cite{CGPR,LPRS}.  However, it can be seen that the effects of the leading perturbative dressing at best give access to the missing information that is exponentially suppressed with distance\cite{SGsub}, and there does not appear to be a controlled approximation in which an asymptotic observer can access internal information, either with Minkowski or AdS asymptotics.  Moreover, this evolution does not appear to significantly alter the trans-horizon entanglement evident in \eqref{matstate}, nor does incorporating the leading semiclassical backreaction arising from $\langle T_{\mu\nu}\rangle$.\footnote{For some investigation of the latter in two-dimensional models, see \cite{GiNe}.}   

This discussion is of course expected to also readily extend to other kinds of matter fields coupled to gravity.  Moreover, treatment of evolution using a hamiltonian of the form  \eqref{matH}, \eqref{Hama} readily generalizes to incorporate interactions between fields, which {\it e.g.} introduce higher-order terms in \eqref{Hama}.    

The evolution can be  described as producing outgoing Hawking quanta at a rate $\calo(1/R)$, for each species of particle.  Correspondingly, the growth in the entanglement or von Neumann entropy of the BH grows also at a rate $\calo(1/R)$, and so grows to a value $\sim S_{BH}=A/4G$  by the time a significant fraction of the initial BH mass has been lost.

\section{Modification of  LQFT/QGR evolution}

The ``missing information" internal to the BH thus appears to persist in the presence of at least the leading-order gravitational corrections that are described by combined evolution via LQFT and QGR.  If so, that returns us to the apparent inconsistency that has been previously explored: either the information is just lost\cite{Hawk-incoh}, ultimately resulting in a quantum description which unacceptably conflicts with energy conservation\cite{BPS}, or a BH remnant remains, apparently inconsistent with basic stability\cite{CaWi,Pres,WABHIP} and other\cite{Susstrouble} constraints.  These problems thus strongly suggest that there must be some  modification to the preceding description of growth of missing information, which is characterized by growing entanglement entropy.

Taking a step back\cite{QFG,QGQF}, if we assume that the theory is fundamentally quantum mechanical, then we need a description of the quantum state and its evolution that is fundamentally unitary.  The question is what form this state and its evolution take, and how those depart from the LQFT/QGR description.  

One feature of the basic kinematic description of many other quantum systems is division of those into quantum subsystems.  As briefly described above, the question of a precise such division in quantum gravity remains.  However, at least at perturbative level an approximate such division appears possible, and moreover there is no obvious indication that perturbative QGR corrections to this approximation are responsible for solving the missing information problem.  In particular, we have outlined the leading LQFT/QGR description of such evolution in the preceding section, and have seen that if the leading gravitational dressing is included, one still has a description of the state with an approximate subsystem division, and consequent buildup of entanglement.
 
Thus, if there is a meaningful approximate division into subsystems, one is led\cite{BHthm} to consider modifications to the evolution of a combined system of a BH and its environment.  One feature of the LQFT/QGR description is that it can be given in different gauges, and so a specific description involves choosing a particular gauge, represented by choice of slicing and spatial coordinates\cite{SE2d}.  One might likewise anticipate that the more fundamental description has analogous gauge freedom, and likewise describe that in a specific gauge.

Given the wide-ranging success of LQFT and GR, extending both into the regime of predictions from early cosmology and now strong gravitational fields of black holes, it seems plausible that the more fundamental description involves departures from these that are small in most circumstances, in particular plausibly including in describing observations of observers falling into a BH.  Thus, we seek to parameterize the more complete dynamics in terms of corrections to LQFT/QGR that for most purposes remain small.

One possible correction to the LQFT/QGR description is the presence of new quantum  interactions between a BH and its surroundings, which are capable of transferring information, yet have limited effect on observers near or falling into the BH.  This is the basic idea of ``nonviolent unitarization."  

Another recent idea\cite{AEHPV} is that there is a new form of evolution of the BH and its surroundings, which acts in a nonunitary (or nonisometric) fashion on the degrees of freedom of the LQFT description, yet preserves ultimate unitarity of scattering.  Ref.~\cite{AEHPV} advocates that this be understood in terms of a map to a ``holographically-dual" theory, on whose Hilbert space the evolution is unitary.  

In either case, we do not yet have complete models, but we do have models capturing some of the important features.  One approach to understanding the basics of such models, and in particular some essential questions of how information flows or transfers, is to work with simplified qubit models.  We next turn to their description, and comparison, before returning to the question of evolution in a more complete model that matches onto the evolution of LQFT in familiar regimes.

\section{Qubit models for evolution}

An approach to describing important features of black hole evolution is to simplify, and describe qubit (or more generally qudit) models for their states. An early such description of Hawking evolution was given in \cite{Mathinforev}, based partly on the simple form of the more complete Hawking state in two dimensions\cite{GiNe}.  Qubit models modifying Hawking's description then have been considered in \cite{Mathinforev,SGmodels,BHQIUE,GiShone} and other subsequent works.

\subsection{Hawking evolution}

Instead of the states $|{\{n_a\}\{n_{\hat a}\}\{n_{\tilde A}\}}\rangle$ of the full $G^0$ Fock space, one can model Hawking evolution in terms of qubit states, at the same time discretizing time evolution in terms of steps, $T=NR$, for integer $N$.  The pairwise emergence of inside and outgoing excitations of  \eqref{stateform}, \eqref{matstate} can be described by the qubit state
\beq\label{HawkQ}
|\psi(T)\rangle = \frac{1}{2^{N/2}}\prod_{i=1}^N \left(|\hat 0\rangle |0\rangle + |\hat 1\rangle |1\rangle\right)_i\ ,
\eeq
which lives in the Hilbert space $\hat\calh_T\otimes \calh_T=\hat \calq^{\otimes N}\otimes \calq^{\otimes N}$, where $\calq$ denotes a basic qubit Hilbert space.  We may include infalling matter by also tensoring in a space $\tilde \calh_T =  \tilde \calh_T \otimes \tilde \calh^{env}_T$.  Here the factorization at time $T$ is that corresponding to describing an excitation as inside the BH, or in its exterior environment, and each of the factors is likewise a product of qubit spaces $\tilde \calq$.  

Description of these states  and their evolution can be facilitated by introducing fermionic annihilation/creation operators $\psi_i$, $\psi_i^\dagger$, satisfying the basic relations
\beq
|1\rangle_i = \psi_i^\dagger |0\rangle_i\quad ,\quad |0\rangle_i = \psi_i |1\rangle_i\quad,\quad \psi_i |0\rangle_i= \psi_i^\dagger |1\rangle_i =0\ ,
\eeq
and likewise for the hatted and tilded spaces.  

A model for the energy-eigenbasis hamiltonian \eqref{Eham} is 
\beq\label{Hzero}
H_0=\sum_i \omega_0 \left(\psi_i^\dagger \psi_i - \hat \psi_i^\dagger \hat \psi_i\right)\ ,
\eeq
with $\omega_0\sim 1/R$ a fixed energy of the Hawking quanta.  This is supplemented by a ``left-moving" hamiltonian $\tilde H_0$ which translates qubits in $\tilde \calh$.  The hamiltonian $H_0$ annihilates the Hawking state \eqref{HawkQ}.  However, the form of the state indicates that at each stage an additional pair of entangled qubits is added.  Alternately, one could, as with \eqref{stateform}, think of those factors as being present throughout the evolution, but inaccessible to low-energy observables until after timestep $N=i$.  

The growing number of accessible qubits leads to growing entanglement of the environment of the BH with its internal states, and this entanglement is part of the description of the missing information.  As the model makes clear, and as holds in more general BH evolution beyond such models, there is a second source of BH entanglement: one may entangle an excitation in $\tilde \calh$ with another outside excitation, and then allow the first to evolve into the interior space via $\tilde H$.  Then, if the BH evaporates and disappears, both types of entanglement are associated to problematic unitarity violation.  We refer to these as type I and type II entanglement, respectively.  (Interactions in more complete models may mix these.)

The large entropy associated with type I entanglement may be illustrated by introducing a new basis of states $|K\rangle_N\in \calq^{\otimes N}$:
\beq
|K\rangle_N =  |0\rangle |0\rangle\cdots  |0\rangle\quad,\quad |0\rangle |0\rangle\cdots  |1\rangle\quad ,\quad\cdots\quad,  \quad |1\rangle |1\rangle\cdots   |1\rangle\ .
\eeq
In terms of these $\caln_N=2^N$ states, and hatted counterparts, the Hawking state may be written 
\beq\label{Hawkexp}
|\psi(T)\rangle = \frac{1}{2^{N/2}}  \sum_{K}^{\caln_N}|\hat K\rangle_N  |K\rangle_N \ .
\eeq
The corresponding von Neumann entropy of the exterior density matrix is $N\ln 2$.

The growth in both types of entanglement during evolution is a central feature of the information problem of Hawking evolution; in order to instead describe unitary evolution, in which the BH is permitted to disappear at the end, the entanglement should begin to decrease as the black hole loses energy, reaching zero at the end -- as in the case of the cavity described in the introduction.  An early description of this behavior appeared in the work\cite{Pageone,Pagetwo}.  We next explore different kinds of models for such reduction in the entanglement.

\subsection{Nonisometric evolution}

Recent work\cite{AEHPV} has given qubit models for BH evolution that are based on a certain form of nonisometric evolution.  In order to better understand the meaning and implications of these models, and their comparison with other kinds of models, we will connect them to the language used in this paper to describe real-time evolution.

The models of \cite{AEHPV} are based on the idea that there are two ``holographically equivalent" descriptions that they call the ``effective" and ``fundamental" descriptions.  Since our focus will be on the description of a ``bulk" observer -- which {\it e.g.} could be a detector falling into a BH, and we will be agnostic about the meaning and origin of a possibly more fundamental description, we will simply describe these in terms of Hilbert spaces $\bar \calh$ and $\calh'$, respectively, referring to the latter as the prime description, and exploring the relation between them.

Specifically, we examine the time evolution of the state between timesteps, as a modification to the Hawking evolution described above.  Before modification, the state at time $T$ is taken to live in the Hilbert space 
\beq
\bar \calh_T = \hat\calh_T \otimes \tilde \calh_T  \otimes \calh_T\otimes \tilde \calh^{env}_T\ ,
\eeq
where again $\hat\calh_T \otimes \tilde \calh_T$ describe the ingoing and outgoing internal states of the BH, respectively, and the evolution was described in the preceding section.

Ref.~\cite{AEHPV} modifies this evolution according to the following description, after translating from their language.  First, at each timestep one qubit is shifted from the space $\tilde \calh^{env}_T$ to join $\tilde \calh_T$. 
Then, a unitary $U_T$ acts on the latter combined space; different scenarios are considered for how strongly this scrambles the states.  At the same time, two new pairs of qubits are appended to 
$\hat\calh_T\otimes \calh_T$,
in the Hawking state 
\beq
\frac{1}{2} \left(|\hat 0\rangle |0\rangle + |\hat 1\rangle |1\rangle\right)_{2N+1}\left(|\hat 0\rangle |0\rangle + |\hat 1\rangle |1\rangle\right)_{2N+2}\ .
\eeq
Finally, a projection operator 
\beq\label{projop}
P_T=2\left( \langle \tilde 0|_{N+1} \langle \hat 0|_{2N+1} +\langle \tilde 1|_{N+1} \langle \hat 1|_{2N+1} \right) \left( \langle \tilde 0|_{N} \langle \hat 0|_{2N+2}+ \langle \tilde 1|_{N} \langle \hat 1|_{2N+2}\right)
\eeq
acts on two qubits of the output of $U_T$ and on the two internal qubits that were added.  The net result is a nonisometric map $M_T$ from the space
\beq
 \tilde \calh_T   \otimes \tilde \calq \otimes \hat \calq^{\otimes 2}  \otimes \calq^{\otimes 2}
\eeq
to the space
\beq\label{outspace}
 \tilde \calh_{T+R} \otimes  \calq^{\otimes 2}\ ,
 \eeq
 where $ \tilde \calh_{T+R} $ has one fewer qubit than $\tilde \calh_T$, and where $\calq^{\otimes 2}$ augments the outgoing radiation space $\calh_T$.  

Ref.~\cite{AEHPV} provide additional interpretation for this by defining the Hilbert space (``fundamental" Hilbert space) $\calh_T'= \tilde \calh_{T+R}$.  
Then, evolution is described as the unitary $U_T$ acting on $\calh_{T-R}'\otimes \tilde \calq$, followed by identifying two of the output qubits as radiation qubits, to map to the space $\calh'_T\otimes \calq^{\otimes 2}$ in \eqref{outspace}.  As part of this interpretation, they define a ``holographic map" $V_T$.  This is related to the preceding by first defining an intermediate ``holographic map" $v_T$ from $\tilde \calh_T   \otimes \tilde \calq \otimes \hat \calq^{\otimes 2}$ to the space $\calh'_{T}$;  $v_T$ arises from combining $U_T$ with the operator $P_T$ of \eqref{projop}.  
Then, $V_T$ is defined iteratively by $V_T=v_T v_{T-1}\cdots v_0$.  

We next compare this evolution with Hawking evolution, to investigate its features.  A first small difference is that two radiation quanta are emitted at each step, while one infaller enters.  
A larger difference is that the evolution does not build up type I entanglement, as a result of the projection $P_T$; it also transfers the type II entanglement out of the BH.  Correspondingly, there is no increase in the dimension of $\hat \calh_T$ at each step.

It does accomplish this at the price of introducing fundamentally nonlocal evolution:  the projection $P_T$ acts in a nonlocal way on the quantum state.  The combined evolution $M_T$  is then both nonlocal and nonunitary on the underlying Hilbert space corresponding to the QFT description.  This forces the type II entanglement out of the BH in a different way than entanglement escapes via interactions in systems like the cavity discussed in the introduction.

To explore this further, note that $M_T$ (and correspondingly the ``holographic map" $v_T$) has null states.  The first factor in $P_T$ annihilates the two states $| \tilde 0\rangle_{N+1} | \hat 1\rangle_{2N+1}$ and $| \tilde 1\rangle_{N+1} | \hat 0\rangle_{2N+1}$, and similarly for the second.  Null states in evolution seemingly present a ``nowhere to go" problem:  a branch of the wavefunction evolves to nothing.  

One might think this might be ameliorated by the fact that it doesn't restrict the infalling matter, but only states of the Hawking quanta, which are not input from the exterior.  Specifically, if the internal evolution $U_T$ outputs $|\tilde 0\rangle_{N+1}$ the projection requires that the outgoing quanta be in the state $|\hat 0\rangle_{2N+1}| 0\rangle_{2N+1}$; if it outputs $|\tilde 1\rangle_{N+1}$, it requires $|\hat 1\rangle_{2N+1}| 1\rangle_{2N+1}$.

This does seem to raise potential issues, however, when extended to a more complete model.  A first one is a potential firewall problem.  Once the Hawking state is projected as in \eqref{projop}, it no longer has the entanglement structure needed for infalling observers to see vacuum-like behavior.  So, if this step in the evolution acts while excitations are near the horizon, it will yield high-energy excitations as seen by those observers.  Possibly, this could be avoided as in nonviolent unitarization, if the evolution $M_T$ acts once the Hawking quanta have reached large, {\it e.g.} $\calo(R)$ wavelengths, but that does also extend the nonlocality and nonunitarity over those scales.

 But beyond that, the internal ``outgoing" modes, modeled by the qubits $\hat \calq$, can participate in interactions in a more complete theory, raising the question of the role of the projections in that context.  A simple example (see \cite{SE2d} for discussion) is decay behind the horizon of a massive infalling particle into lighter particles, one of which is inward going and one of which is outward going, which will then populate the modes corresponding to $\hat \calq$.  Then, there is the question of how to describe consistent evolution, in view of the ``nowhere to go" problem.  Or, more picturesquely, an infalling observer could fire their rocket engines after crossing the horizon, in a vain attempt to escape.  They would not do so, but could become an internal outward mover.  This again raises the question of a consistent extension of the projections \eqref{projop}, acting on outward movers.
Additional questions raised by interactions, {\it e.g.} regarding ultimate unitarity, have also been recently considered in \cite{KiPr}.

\subsection{Nonviolent unitarization}

The approach of nonviolent unitarization\cite{SGmodels,NVNL,NVUEFT,NVNLT}\cite{NVU,BHQU} (NVU)  considers interactions between the BH Hilbert space and the environment, which can transfer information or entanglement out of the BH, but in a way that `minimally' affects the standard description of BHs, {\it e.g.} as described by infalling or other exterior observers.  Accomplishing the information transfer via interactions is similar to the physics by which the cavity described in the introduction reduces its entanglement.

Qubit models for NVU are found by augmenting the hamiltonian \eqref{Hzero}: $H=H_0+\Delta H$, with two additional pieces.  The first is a term $\Delta H_I$ which describes the interactions between BH and environment states, and so can transfer entanglement from BH to environment.  The second is a term $\Delta H_{bh}$ which alters the evolution of the internal state of the BH.  The internal part of the state \eqref{HawkQ} has a ``frozen" character, in that the internal excitations simply persist unchanged; similar behavior is found for more complete states \eqref{matstate} based on nice slicings that avoid the singularity\cite{SEHS,SE2d}\cite{GiPe1}.  This is a feature of a particular gauge or picture for describing the state, which artificially retards evolution into the region near $r=0$.  In a more complete quantum description of the interior evolution, it is quite plausible that the natural dynamics ultimately leads to nontrivial evolution; {\it e.g.} it might begin mixing or scrambling the internal state on some timescale.  $\Delta H_{bh}$ is introduced to describe this aspect of evolution, which is analogous to mixing of internal states in the cavity described in the introduction.

At the level of detailed qubit models, there are multiple possibilities.  $\Delta H_{bh}$ could induce general transitions  between BH states $|\hat K\rangle$, as well as mixing with the infalling states 
 $|\tilde K\rangle$.  A particularly simple possibility is if $\Delta H_{bh}$ conserves $H_0$.  It could couple multiple internal qubits, or even have a form as simple as $\Delta H_{bh}=h_{ij} \hat \psi^\dagger_i \hat \psi_j$ and/or $\Delta H_{bh}=h_{ij} \hat \psi_i \tilde \psi_j$, rearranging qubits.  We will also assume that there is a ``latency time" $\Delta T= R\Delta N$, such that $\Delta H_{bh}(T)$ only acts nontrivially on the first $N-\Delta N$ qubits; before the latency time, the internal qubits in \eqref{HawkQ} are unscrambled.  Different forms for $\Delta H_{bh}$ may also follow from different choices of gauge to describe internal evolution.  

The transfer term $\Delta H_I$ -- which would be forbidden by locality in a QFT description -- plays the key role of transferring entanglement to the environment, and the structure of this term is particularly interesting since it bears on how the BH appears from the outside.  The simplest form for such a term is
\beq
\Delta H_I \sim \gamma_{\hat \alpha\alpha}  \hat \calo_{\hat \alpha} \calo_\alpha
\eeq
with some coefficients $\gamma_{\hat \alpha\alpha}$, and with general operators acting on the internal excitations and external excitations, respectively; if one includes infalling matter, there could likewise be terms with internal operators $\tilde\calo_{\tilde \alpha}$.  These two kinds of terms can then transfer type I or type II entanglement, respectively, to the BH's environment.

For a  simple model, consider the case where $\Delta H_I$ couples to a single qubit in the interior, $\hat \calo = \hat \psi_i^\dagger$.  Recall from \eqref{Hzero} that $\psi_i^\dagger$ lowers the BH energy; a particularly simple model would have interactions $\gamma_{ij} \hat \psi_i^\dagger \psi_j^\dagger$, conserving energy, although we will also consider coupling to a more general external operator.  We can also generalize these interactions to couple to the operators $\tilde \psi_i$ in the interior.  

To model the possibly delayed turn on of such interactions, append a qubit $\calq_0$ to the qubits of the Hawking state \eqref{HawkQ}, and consider concretely  
\beq\label{HIex}
\Delta H_I = \gamma \hat \psi_0^\dagger \calo + h.c.
\eeq
for some exterior operator $\calo$.  Then, assume that $\Delta H_{bh}$ acts to mix $\calq_0$ and  qubits with $i=1,\cdots N-\Delta N$, including the latency time  parameter $\Delta T$.  
After this time, the qubits of the Hawking state \eqref{HawkQ}, mix with $\calq_0$; the interaction \eqref{HIex}  is then sensitive to their state, and can transfer their information into the state of the environment via the exterior operator $\calo$.

Key questions, then, are whether $\Delta H_I$ transfers information at a sufficient rate to address the unitarity problem, what effects it has on dynamics in the vicinity of the BH, and what effect dynamics  has on observers who have fallen in.

Estimating the rate at which \eqref{HIex} induces transitions gives an estimate of the rate at which it transfers information (more precisely, entanglement)\cite{NVU,GiRo}.  Consider its action on a state which is a general (for simplicity) $N$-qubit state $|\psi_{bh}\rangle$ of the interior, and $|\psi_e\rangle$, which could be vacuum, for the exterior.  Transition rates to different final states can be estimated from the matrix elements
\beq\label{matel}
\gamma  \langle \psi_{bh}'| \psi_0^\dagger |\psi_{bh}\rangle    \langle\psi_e'|   \calo |\psi_e\rangle\ 
\eeq
of $\Delta H_I$.

Suppose that $|\psi_{bh}\rangle=\sum_K c_K |\hat K\rangle$ is a generic unit-normalized BH state, arising from the evolution with $H_0+\Delta H$ acting on the Hawking state \eqref{Hawkexp}, and projecting on one of the internal components.  Then, the matrix element to a generic final state is of the same size as the generic sizes of the $c_K$,
\beq
\langle\psi_{bh}'| \psi_0^\dagger |\psi_{bh}\rangle \sim \frac{1}{2^{N/2}}=e^{-S_{bh}/2}\ ,
\eeq
Here we have defined a black hole entropy by $\exp\{S_{bh}\}=2^N$, for comparison with other discussions.  While this effectively weak coupling to a generic state suggests  a tiny transition rate, there are $\calo(2^N)=e^{S_{bh}}$ final states which one can transition to.  So, an argument based on Fermi's Golden rule\cite{NVU} suggests that if the matrix element of the operator $\gamma \calo$ in \eqref{matel} is $\calo(1/R)$, the total transition rate, and hence the information transfer rate, will be the same size, $\calo(1/R)$.  This, in order of magnitude, is the needed size to compensate for the growth of type I entanglement due to Hawking evolution, and in a more general scenario, can also transfer any type II entanglement that has been injected into the BH.

This rate can be explained from a different argument.  For generic qubit state $|\psi_{bh}\rangle$, $\psi_0^\dagger$ acts to create {\it some} state with $\calo(1)$ norm.  Then, the transition rate to that state will not be exponentially suppressed.

This kind of argument -- which can be refined\footnote{For {\it some} further discussion, see \cite{GiRo}.} -- sets the size of the couplings to the operators $\calo$, which we think of as acting on a ``quantum atmosphere" of the BH.  The same methods can be used to estimate the amplitude if the initial external state $|\psi_e\rangle$ is not vacuum, but instead something else, {\it e.g.} an incoming gravitational wave state\cite{BHQU}; that just changes which matrix element between external states enters in \eqref{matel}.  In either case, the ``nonviolence" assumption\cite{SGmodels,NVNL,NVUEFT,NVNLT}\cite{NVU,BHQU} is that the operator $\calo$ only transfers soft momenta and energies to the external state, {\it e.g.} with $\calo(1/R)$ sizes, although more general powers of $R$ can also be considered.  In qubit models, this is naturally implemented since the qubit excitations have energies $\sim 1/R$, as long as $\calo$ only acts on qubits that have begun to exit the horizon region.  The final result is that the interactions \eqref{HIex} not only yield entanglement transfer rates of size $\calo(1/R)$, but also 
can give $\calo(1)$ modifications to scattering amplitudes which have energy and momentum transfers $\calo(1/R)$.  

One can ask about other effects besides corrections to these amplitudes, {\it e.g.} on an infalling external observer.  From this outside perspective, a simple test is to inquire whether  \eqref{HIex} can give a significant coherent shift to the external hamiltonian.  To estimate the size of such a shift, consider the expectation value in a generic BH state,
\beq\label{HIexp}
\Delta H_I^{eff} = \langle \psi_{bh}|\hat \psi_0^\dagger| \psi_{bh}\rangle \gamma \calo + h.c.\ .
\eeq
The size can be estimated by expanding 
\beq
|\psi_{bh}\rangle = |\hat0\rangle_0 |\phi_1\rangle +  |\hat1\rangle_0  |\phi_2\rangle\ ,
\eeq
where for a generic state $|\psi_{bh}\rangle$, the states $ |\phi_1\rangle$, $ |\phi_2\rangle$ are generic states of the qubits excluding $\hat \calq_0$. Then the expectation value in \eqref{HIexp} is $\langle \phi_2|\phi_1\rangle$, which is also of order $2^{-N/2} = e^{-S_{bh}/2}$, and so such a coherent effect is exponentially small, as also found in \cite{NVU}.

Another question is that of the effect on internal observers.  Note that the hamiltonian \eqref{HIex}, or its generalization to the case where  $\Delta H_{bh}$ also mixes  the infalling modes in $\tilde \calh$, has {\it no} effect on internal excitations until after the latency time $\Delta T$.  This could be very long as compared to the expected infall time, $\sim R$, of an infalling observer, and as noted may only become important at much longer times where deviations from nice slice evolution become important.  Once internal excitations do become mixed with the excitations on which $\Delta H_I$ acts (in the simple model, just $\hat \calq_0$), those excitations are non-trivially acted on, corresponding to, {\it e.g.}, particle creation or annihilation in the relevant modes.

Finally, we have not discussed the second term in \eqref{HIex}, arising from the hermitian conjugate.  Note that if the first term can lower the energy of the BH and produce an excitation in the atmosphere, this term has the opposite effect.  However, we can compare this situation to that of the cavity.  There, the interaction term between internal and external excitations, arising from the presence of the small hole, has the same nature: it can either transfer particles (and entanglement) out of the cavity, or back into the cavity.  The reason that transfer out dominates is due to phase space considerations:  there are many more states outside the cavity than inside, and when one accounts for the full hamiltonian, excitations will dominantly fill those external states after sufficiently long time.  We expect that a more complete dynamical model should exhibit such behavior.

\section{Comparing models, and towards a more complete description}

We conclude with further discussion of the comparison of the two types of models, and  of possible features of their generalizations to a more complete evolutionary law, modifying the LQFT/QGR evolution described in Sec.~\ref{QFTsec}.  

In the qubit models, the nonisometric evolution of \cite{AEHPV} has been seen to combine unitary evolution of the internal state of the BH with a fundamentally new kind of evolution, which involves projections correlating the quanta of the would-be Hawking state with other internal excitations.  This evolution accomplishes transfer of entanglement from the BH internal state, but thus in a way that is different from the quantum dynamics of other known fundamental systems with subsystems, like the cavity example described in the introduction. 
These operations are neither local nor unitary, when acting on the degrees of freedom  of the familiar LQFT description, and in particular nonlocally transfer entanglement to degrees of freedom near the horizon.  
Ref.~\cite{AEHPV} suggests that the evolution be reinterpreted as unitary evolution in a holographically-dual fundamental evolution.   

In particular, the projection eliminates certain states of the outgoing internal excitations, which, if excited, then have ``nowhere to go" in the evolution, and this also prevents buildup of type I entanglement due to the Hawking evolution.  In the simplest models, only the Hawking evolution produces such excitations.  However, in a more realistic model interactions can also populate these modes, raising the question of how to describe a more complete consistent evolution incorporating the nonisometric projections.  A very simple example\cite{SE2d} is decay of an infalling particle, producing decay products in outgoing internal modes.  Other questions regarding interactions are explored in \cite{KiPr}.

The models also eliminate excitations from the internal matter, through the projections; for example, after the projection \eqref{projop}, the two qubits $\tilde \calq_N$ and $\tilde \calq_{N+1}$ 
are no longer part of the internal system, and so one likewise expects ``disappearance" of excitations or particles from the internal state in a more complete model, having an effect on internal observers.  The models do however offer the apparent virtue of hiding this elimination of excitations, due to the complexity of the unitary internal mixing\cite{AEHPV}.

If the Hilbert space structure of quantum gravity is such that the BH and environment are distinct subsystems, and evolution both describes ultimate BH disappearance and is instead unitary, one requires\cite{BHthm} interactions to transfer the information to the external state.  Scenarios for this have been investigated, with the goal of describing interactions with ``minimal" effect on infalling or other near-BH observers, in the approach of 
``nonviolent unitarization"\cite{SGmodels,NVNL,NVUEFT,NVNLT}\cite{NVU,BHQU};
such interactions have a different effect on the respective subsystems than the nonisometric models.  
Models for nonviolent unitarization that go beyond the qubit models surmount some important challenges, although there are also remaining questions regarding a more complete dynamics.  

As with the nonisometric models, one does pay the price of apparent nonlocality with respect to the semiclassical background BH geometry.  However, the interactions of nonviolent unitarization, while they can give order one corrections to amplitudes, only do so for situations with soft momentum and energy transfer, for example with characteristic wavenumber or energy $1/R$.  This has nearly negligible effect on small observers falling into a BH; put differently, the departures from the equivalence principle are likewise extremely modest, and vanish in the large-$R$ limit.

Such interactions do not have the same kind of puzzling effects on the internal outgoing excitations as the nonisometric evolution.  However, through the combination of modified internal evolution due to $\Delta H_{bh}$ and the interactions $\Delta H_I$, this evolution is expected to have effects on  both ingoing and outgoing inside modes that depart from evolution described by LQFT, and which, {\it e.g.}, lead to disappearance of excitations.  Depending on the details of the model, this may only affect inside excitations after long times.  One motivation for this is if the departure from the nice-slice hamiltonian and evolution of LQFT is a small effect, and so only builds up after long-time evolution, where the nice slice gauge for early excitations becomes increasingly extreme.
This evolution moreover could share the feature of hiding the effects of information transfer\cite{AEHPV} on inside excitations if the internal evolution via $\Delta H_{bh}$ is sufficiently complex.   

One important question in the models we have described is whether the entanglement transfer from the BH is complete by the time that the BH disappears; otherwise one has a potentially problematic remnant.  The simple evolution described above, with combined scrambling of excitations and transfer via interactions, is analogous to the cavity model described in the introduction.  However, as a BH nears the end of its evolution, its rate of energy loss accelerates, which is different from the slowdown of information transfer expected in such a cavity model.  One expects that a more complete hamiltonian would describe an effect analogous to squeezing of the cavity to accelerate information transfer.  One effect that could at least partly account for this is an increase in the internal energy levels with decreasing BH size.  Further aspects of a more complete description of evolution are to be explored in future work.

\vskip.3in
\noindent{\bf Acknowledgements} 

I thank D. Harlow, S. Mathur and J. Perkins for useful discussions.  This material is based upon work supported in part by the U.S. Department of Energy, Office of Science, under Award Number {DE-SC}0011702, and by Heising-Simons Foundation grant \#2021-2819.   This work was initiated at the Aspen Center for Physics, which is supported by National Science Foundation grant PHY-1607611.

\mciteSetMidEndSepPunct{}{\ifmciteBstWouldAddEndPunct.\else\fi}{\relax}
\bibliographystyle{utphys}
\bibliography{models}{}

\end{document}